\documentstyle[12pt]{article}

\newcommand{\bm}[1]{\mbox{\boldmath $#1$}}
\newcommand{\be}{\begin{equation}}
\newcommand{\ee}{\end{equation}}
\newcommand{\bea}{\begin{eqnarray}}
\newcommand{\eea}{\end{eqnarray}}
\newcommand{\nn}{\nonumber}

\makeatletter
\def\section{\@startsection{section}{1}{\z@}{-3.5ex plus -1ex minus -.2ex}{2.3ex plus .2ex}{\normalsize \bf}}
\def\subsection{\@startsection{subsection}{2}{\z@}{-3.25ex plus -1ex minus -.2ex}{1.5ex plus .2ex}{\normalsize \sl}}
\def\subsubsection{\@startsection{subsubsection}{3}{\z@}{-3.25ex plus -1ex minus -.2ex}{1.5ex plus .2ex}{\normalsize}}
\makeatother

\makeatletter
\def\@evenhead{
	\vbox{\hbox to\hsize{\bf \thepage \hfill \sl \evenmark}
	}
}
\def\@oddhead{
	\vbox{
		\hbox to\hsize{\oddmarkA \oddmarkB \hfill \oddmarkC}
		\hbox to\hsize{\hfill \oddmarkD}
		\vspace{.03in}
		\hbox to\hsize{\hfill \oddmarkE}
		\vspace{.03in}
		\hbox to\hsize{\hfill \oddmarkF}
	}
}
\def\@evenfoot{\vbox{\hbox to\hsize{\hrulefill}
	\vspace{-40pt} \hbox to \hsize{\today \sl \footmsg \hfill}
	}
}
\def\@oddfoot{
	\vspace{-40pt} 
	\hbox to \hsize{ \footmsgA \hfill }
}
\makeatother

\def\therevisionhistory{Revision history: April 1996. Presented Bayesian Statistics 6, Valencia, 1998. Invited paper Monograph on Bayesian Methods in the Sciences, Rev. R. Acad. Sci. Exacta. Fisica. Nat. Vol. 93, No. 3, pp. 381--386, 1999. Arxiv version asserts bold vectors dropped in print.}

\def\evenmark{\bf entropy}
\def\oddmarkA{{\sl Maximally Informative Statistics}}
\def\oddmarkB{\thepagerange}
\def\oddmarkC{}
\def\oddmarkD{}
\def\oddmarkE{}
\def\oddmarkF{}
\def\footmsgA{{\thecopyright}}

\def\thedoctitle{\bf Maximally Informative Statistics}

\def\firstauthorname{David R. Wolf}
\def\secondauthorname{Edward I. George}

\def\firstauthoraddress{PO 8308, Austin, TX 78713-8308, USA, E-mail: drwolf@realtime.net \\  Dr. Wolf is the corresponding author. \\}

\def\secondauthoraddress{Department of MSIS, University of Texas,
      Austin, TX 78712-1175, Email: egeorge@mail.utexas.edu \\ }

\begin{document}

%%% move down
{
	\noindent {\\ \\}
}

%%% format the title
{
	\LARGE \noindent \thedoctitle
} \\

%%% format first author name
{
	\noindent {\bf {\firstauthorname }
} \\

%%% format first author address
{
	\normalsize \noindent {\firstauthoraddress}
} \\

%%% format second author name
{
	\noindent {\bf {\secondauthorname}}
} \\

%%% format second author address
{
	\normalsize \noindent {\secondauthoraddress}
} \\

%%% format revision history
{
	\noindent {\sl \therevisionhistory}
} \\

\vspace{2pt} \hbox to \hsize{\hrulefill}
\vspace{.1in}

%\begin{inmargins}{1cm}
\noindent
{\bf Abstract:} In this paper we propose a Bayesian, information theoretic approach to dimensionality reduction. The approach is formulated as a variational principle on mutual information, and seamlessly addresses the notions of sufficiency, relevance, and representation.  Maximally informative statistics are shown to minimize a Kullback-Leibler distance between posterior distributions.  Illustrating the approach, we derive the maximally informative one dimensional statistic for a random sample from the Cauchy distribution. \\


\vspace{2pt} \hbox to \hsize{\hrulefill}

\newpage

%%% reset the headers and footers for rest of pages

\def\oddmarkC{\thepage}
\def\oddmarkB{}
\def\oddmarkD{}
\def\oddmarkE{}
\def\oddmarkF{}
\def\footmsgA{}

%%% sections

\section{Introduction}

Dimensionality reduction is a fundamental goal of statistical science.   In a modeling context, this is often facilitated by estimating a low dimensional quantity of interest.  For example, suppose the quantities of interest are the labels of a classification of photographs of objects; of trees, children, etc.  The data are the photographs, and the goal is to infer which of the several classes have been presented.  In this case the data space often has dimension on the order of $>10^{6}$, while the parameter space is a small discrete set of labels each having much lower dimension.   A low dimensional summary of the photograph is then obtained as the estimate of the classification of the photograph.

In this paper, we propose a novel fully Bayesian information theoretic approach to dimensionality reduction, based on maximizing the mutual information between a statistic and a quantity of interest.  The approach is formulated as a variational principle on mutual information, and it seamlessly addresses the notions of sufficiency, relevance, and representation.  We refer to statistics which maximize this mutual information as {\em maximally informative} (MI) statistics.  Such statistics are shown to minimize a Kullback-Leibler distance between posterior distributions.

The mutual information between a statistic and  a quantity of interest is defined in section~\ref{sec:mutinfo}.  The mutual information based variational principle for MI statistics is utilized in section~\ref{sec:maxinform} to derive non-variational derivative forms of the principle. In section~\ref{sec:suff} several properties of MI statistics are derived.  The important result of this section is that MI statistics provide a generalization of the notion of sufficiency,  because they are sensible both when they are not sufficient statistics, and when lower-than-data-dimension sufficient statistics do not exist. In section~\ref{sec:kull} we present the result that in inference the Kullback-Leibler (KL) distance is properly a functional of posterior distributions. There we find MI statistics at functional minima of a KL distance based on posterior distributions of the parameter of interest.  The arguments made here suggest that the KL distance derived here is to be preferred to a maximum relative entropy distance, a fact which is not discussed in, for example, Kullback \cite{KULL59} or Shor \cite{SHOR79}, and numerous others.  In section~\ref{sec:Gauss} the MI static for the location parameter of the univariate Gaussian distribution is derived, and shown to be the expected result, since in this case a one-dimensional sufficient statistic exists. In section~\ref{sec:Cauchy} we find a one-dimensional MI statistic for the Cauchy distribution, where a sufficiency reduction does not exist.  In section~\ref{sec:ApproxAndBayes} we discuss approximating the posterior distribution as a Gaussian and apply this technique to show that the MI statistics are then Bayes' estimators of the mean and standard deviation.  In that section a contrast of the approximate MI inference approach with the Maximum Entropy method is made, and it is shown that although they agree for Gaussian likelihoods, they disagree for other distributions, with simplicity arguing in favor of the MI statistics.

\section{The Mutual Information Between a Statistic and a Quantity of Interest
 \label{sec:mutinfo}}

Let the data $\bm{x} \in X$ be drawn according to a parameterized distribution $P(\bm{x} \mid \bm{\theta})$, with $\bm{\theta} \in \Theta$, the parameter space.   $\bm{\theta}$ itself is distributed according to the prior $P(\bm{\theta})$.  The marginal distribution of $\bm{x}$ is obtained from $P({\bm{x}}) = \int {P(\bm{x} \mid \bm{\theta}) \, P(\bm{\theta})} \, d{\bm{\theta}}$, and the posterior of $\bm{\theta}$ given $\bm{x}$ is obtained from Bayes Theorem as
\be
P(\bm{\theta} \mid \bm{x}) = \frac{P(\bm{x} \mid \bm{\theta}) \, P(\bm{\theta})}{P({\bm{x}})}
\label{eq:posterior}
\ee

The quantity of interest $\bm{q} = {\bm{\xi}_{Q}}(\bm{\theta})$ will be a function of $\bm{\theta}$, a mapping from the parameter space $\Theta$ into some $Q$, $\bm{\xi}_{Q}(\cdot) : \Theta \rightarrow Q$.  It will be useful to use the Dirac delta-function $\delta(\cdot)$ to represent the distribution of $\bm{q}$ as 
\bea
P(\bm{q} \mid \bm{\theta}) &:=& P( \{\bm{\theta} : \bm{q} = {\bm{\xi}_{Q}}(\bm{\theta}) \} \mid \bm{\theta}) \nn \\
&=&\delta( \bm{q}- {\bm{\xi}_{Q}}(\bm{\theta}) ) \label{eq:deltadef} \\
&=& \Pi_{i=1}^{k_{q}} \delta( {q}_{i}- {\xi_{Q,i}}(\bm{\theta}) ),
\label{eq:distOfq}
\eea
where $\delta(\bm{z}(\cdot)) = \Pi_{i} \delta(z_i(\cdot))$.  
Note that (\ref{eq:deltadef}) may be seen directly by using Bayes' theorem to expand $P(\bm{q}, \bm{\theta})$ as $P(\bm{q} \mid \bm{\theta}) \,  P(\bm{\theta})$, integrating that over $\bm{q}$, which must produce $P(\bm{\theta})$, and noting that because the support of $P(\bm{q} \mid \bm{\theta})$ is the unique $\bm{q}$ such that $\bm{q} = {\bm{\xi}_{Q}}(\bm{\theta})$ ($\bm{\theta}$ is specified), $P(\bm{q} \mid \bm{\theta})$ must therefore be the Dirac delta function. 
The distribution of $\bm{q}$ given the data $\bm{x}$,
may be written using (\ref{eq:posterior}) and (\ref{eq:distOfq}) as
\bea
P(\bm{q} \mid \bm{x})
&=& \int P(\bm{q} \mid \bm{\theta} ) \, P({\bm{\theta}} \mid {\bm{x}}) \, d{\bm{\theta}}
\eea

A statistic $\bm{r} = \bm{\xi}_{R}(\bm{x})$ will be a function of $\bm{x}$, a mapping from the data space $X$ into some $R$, $\bm{\xi}_{R}(\cdot): X \rightarrow R$.  Again using the delta notation, the distribution of the statistic given data is
\bea
P(\bm{r} \mid \bm{x}) &=& \delta( \bm{r}- {\bm{\xi}_{R}}(\bm{x}) ) \\ 
&=& \Pi_{i=1}^{k_{r}} \delta( r_{i}- {{\xi}_{R,i}}(\bm{x}) )
\label{eq:indOfData}
\eea
The joint distribution of the statistic $\bm{r}$ and the quantity of interest $\bm{q}$, conditioned on the data $\bm{x}$ is
\be
P(\bm{r}, \bm{q}  \mid \bm{x}) = P(\bm{r} \mid \bm{x}) \, P(\bm{q} \mid \bm{x}) \label{eq:jointGivenData}
\ee
(since $\bm{r} = \bm{\xi}_{R}(\bm{x})$ is specified once $\bm{x}$ is known, making $P( \bm{r} \mid \bm{x}, \bm{q} ) = P( \bm{r} \mid \bm{x} )$), and the unconditional joint distribution is
\be
P(\bm{r}, \bm{q}) = \int P(\bm{r} \mid \bm{x}) \, P(\bm{q} \mid \bm{x}) P(\bm{x}) d\bm{x}. \label{eq:joint}
\ee

Finally, we define the mutual information between a statistic $\bm{\xi}_{R}(\cdot)$ and a quantity of interest $\bm{\xi}_{Q}(\cdot)$ as
\be
M(\bm{\xi}_{R}(\cdot),\bm{\xi}_{Q}(\cdot)) = \int \int P(\bm{r}, \bm{q}) \, \, log\left( \frac{P(\bm{r}, \bm{q})}{P(\bm{r}) \, P(\bm{q})} \right) d\bm{q} \, d\bm{r} \label{eq:mutInf}
\ee
This mutual information is the Kullback-Leibler distance between the joint distribution $P(\bm{r}, \bm{q})$ and the marginal product $P(\bm{r}) P(\bm{q})$ corresponding to independence between $\bm{r}$ and $\bm{q}$.
Note that this Kullback-Leibler distance is different from the Kullback-Leibler distance mentioned in the introduction (and seen later in section~\ref{sec:kull}).  A major contribution of this paper is the demonstration of how these two Kullback-Leibler distances are related.

\section{MI Statistics and the Variational Principle \label{sec:maxinform}}

We now define the maximally informative (MI) statistic.

\begin{quote}
{\em Let $S=\{\bm{\xi}_{R}(\cdot)\}$ be a set of statistics under consideration. A MI statistic for a quantity of interest ${\bm{\xi}_{Q}}(\cdot)$ is any statistic $\bm{\xi}_{R}(\cdot)$ from $S$ maximizing the mutual information $M(\bm{\xi}_{R}(\cdot),{\bm{\xi}_{Q}}(\cdot))$ between the statistic and the quantity of interest.}
\end{quote}

The following variational principle can be used to obtain an MI statistic.  Let $\frac{\delta}{\delta f(\cdot)}$ denote the functional derivative with respect to $f(\cdot)$.

\begin{quote}
{\em Choose $\bm{\xi}_{R}(\cdot)$ from $S$ such that $\frac{\delta M(\bm{\xi}_{R}(\cdot),{\bm{\xi}_{Q}}(\cdot))}{\delta {\bm{\xi}_{R}}(\cdot)} = 0$ and $\frac{\delta^2 M(\bm{\xi}_{R}(\cdot),{\bm{\xi}_{Q}(\cdot)})}{\delta {\bm{\xi}_{R}(\cdot)}^2}$ is negative semidefinite, i.e. so that $\bm{\xi}_{R}(\cdot)$ maximizes the information between itself and $\bm{\xi}_{Q}(\cdot)$, the quantity of interest. If possible, choose the global maximum. }
\end{quote}

Note that MI statistics in $S$ may occur on the boundary of $S$.  This may be a case of interest, which occurs when constraints are imposed on the statistics, and may be handled with a trivial modification.  Note also that the space $S$ of statistics may be constrained to contain only low-dimensional statistics, in order to force a dimesionality reduction of the data. 

We now demonstrate the variational principle for MI statistics.  
The argument proceeds by varying (see, for example \cite{ARFK85} for the variational calculus) the mutual information of (\ref{eq:mutInf}) with respect to the statistic function $\bm{\xi}_{R}(\cdot)$ of dimension $k_{r}$, i.e. $\bm{\xi}_{R}(\cdot) = (\xi_{r,1}(\cdot), \ldots, \xi_{r,k_{r}}(\cdot))$.  We now proceed to substitute $\bm{\xi}_{R}(\bm{x})=\bm{\xi}_{R}^{0}(\bm{x})+\epsilon {\bm{\eta}}(\bm{x})$ in (\ref{eq:mutInf}), and take the derivative with respect to $\epsilon$.  

Assuming appropriate regularity conditions, we have
\bea
\partial_{\epsilon} M(\bm{\xi}_{R}(\cdot),\bm{\xi}_{Q}(\cdot)) &=& \int \int \left[ \partial_{\epsilon} P(\bm{r}, \bm{q}) log \left( \frac{P(\bm{r}, \bm{q})}{P(\bm{r}) P(\bm{q})} \right) \right. \nn \\ 
& & \quad \quad \quad \quad + \left. \rule{0in}{3.0ex} P(\bm{r}) \partial_{\epsilon} P(\bm{q} \mid \bm{r}) \right] d{\bm{q}} d\bm{r} \label{eq:derivMutInf1} \\
&=& \int \int  \partial_{\epsilon} P(\bm{r}, \bm{q}) log \left( \frac{P(\bm{r}, \bm{q})}{P(\bm{r}) P(\bm{q})} \right) d{\bm{q}} d\bm{r}, \label{eq:derivMutInf2}
\eea
where simplification from (\ref{eq:derivMutInf1}) to (\ref{eq:derivMutInf2}) occurs because probability is conserved.  Utilizing (\ref{eq:jointGivenData}) we find
\be
P(\bm{r}, \bm{q})=\int \delta(\bm{r}-\bm{\xi}_{R}(\bm{x})) \, P( \bm{q} \mid \bm{x}) \, P(\bm{x}) \, d\bm{x} \label{eq:jointSubs}
\ee
Taking the derivative of (\ref{eq:jointSubs}) with respect to $\epsilon$ yields
\be
\partial_{\epsilon} P(\bm{r}, \bm{q}) = \Sigma_{j=1}^{k_{r}} \int \delta'(r_{j}-{\xi}_{R,j}(\bm{x}))  \eta_{j}(\bm{x}) \Pi_{i \ne j} \delta(r_{i}-{\xi}_{R,i}(\bm{x})) P( \bm{q} \mid \bm{x}) P(\bm{x}) \, d\bm{x} \label{eq:derivJoint}
\ee
Note that because $\bm{\eta}$ is arbitrary, we may choose it to simplify as needed. 

We proceed by considering $k_{r}$ choices of $\bm{\eta}$.  Label the choices by $m\in \{1, \ldots, k_{r} \}$, and on choice $m$ take the components of $\bm{\eta}$ as follows:
\bea
\eta_{\ell}(\bm{x}) &=& \delta(\bm{x}-\bm{x}_{c}),  \, \, \, \, (\ell = m) \\
\eta_{\ell}(\bm{x}) & =& 0, \, \, \, \, (\ell \ne m)
\eea
where $\bm{x}_{c}$ is any data point we may choose.  The condition that the mutual information is extremal then becomes the statement that for all $\bm{x}_{c}$ and $i \in \{1, \ldots, k_{r}\}$.
\bea
\partial_{\epsilon} M(\bm{\xi}_{R}(\cdot),\bm{\xi}_{Q}(\cdot)) \mid_{\epsilon=0} &=& 0 \\
&=& \int \int  \delta'(r_{i}-\xi_{R,i}^{0}(\bm{x}_{c})) \, \Pi_{i \ne j} \delta(r_{i}-\xi_{R,i}^{0}(\bm{x}_{c})) \,  \nn \\
& & \quad \quad \times P( \bm{q} \mid \bm{x}_{c}) \, log \left( \frac{P(\bm{r}, \bm{q})}{P(\bm{r}) P(\bm{q})} \right) d{\bm{q}} d\bm{r} \label{eq:derivMutInf3} 
\eea

Integrating (\ref{eq:derivMutInf3}) by parts with respect to $\bm{r}$ (dropping both the ``$0$'' superscript and subscript ``$c$'', since there is no distinction to be made at this point) yields the condition that for all $\bm{x}$ 
\be
\int P(\bm{q} \mid \bm{x}) \, \partial_{\bm{r}} log \left( \frac{P(\bm{r}, \bm{q})}{P(\bm{r}) P(\bm{q})} \right) \mid_{\bm{r} = {\bm{\xi}_{R}}(\bm{x})} d{\bm{q}} = 0 \label{eq:derivCondx1}
\ee
where derivatives with respect to vectors are gradients (vectors of derivatives).  The form from which the theorems of the next section are proven, is found by rewriting (\ref{eq:derivCondx1}) as
\be
\int \frac{P(\bm{q} \mid \bm{x})}{P(\bm{q} \mid {\bm{r}})} \, \partial_{\bm{r}} P(\bm{q} \mid \bm{r}) \mid_{\bm{r} = {\bm{\xi}_{R}}(\bm{x})} d{\bm{q}} = 0 \label{eq:derivCondx2}
\ee

\section{MI Statistics and Sufficiency \label{sec:suff}}

Now we prove several important properties concerning MI statistics.  The first property is the intuitively obvious property that {\em data is a MI statistic}. The second property is that {\em any sufficient statistic is a MI statistic}. Finally, we note that {\em MI statistics are not necessarily sufficient statistics}. 

\begin{quote}
{\bf Theorem 1 \label{thm:1}}. {\em Any 1--1 function of data is a MI statistic of the quantity of interest}

Proof: Let ${\bm{\xi}_{R}}(\cdot)$ be the identity so that ${\bm{\xi}_{R}}(\bm{x}) = \bm{x}$ in (\ref{eq:derivCondx2}). The fraction in that equation is then $1$, and the derivative integrates to zero because probability is conserved. Having ${\bm{\xi}_{R}}(\bm{x})$ any invertible function  changes nothing as any value of it determines $\bm{x}$. 
\end{quote}

\begin{quote}
{\bf Theorem 2 \label{thm:2}}. {\em Any sufficient statistic for the quantity of interest is a MI statistic of the quantity of interest}

Proof: Note that, using the definition of ${\bm{\xi}_{R}}(\bm{x})$ being a sufficient statistic, the ratio in (\ref{eq:derivCondx2}) is one - the posterior distribution of the quantity of interest given the data $\bm{x}$ is the same as the posterior distribution of the quantity of interest given the sufficient statistic ${\bm{\xi}_{R}}(\bm{x})$. The derivative then integrates to zero because probability is conserved. 
\end{quote}

(Note that in both Theorems $1$ and $2$ the Hessian condition of the MI inference variational principle is easily established since then the extremum of the mutual information is easily seen to be a local maximum. Otherwise, one must check convexity.)

Although it is true that any sufficient statistic is a MI statistic, the converse is false.  In problems (of data dimension greater than one) where a lower-than-data dimension sufficient statistic does not exist, there will exist a lower-than-data dimension statistic which is MI but not sufficient.  Thus, the class of maximally informative statistics contains the sufficient statistics, but is broader.  MI statistics need not provide all of the available information about the underlying quantity of interest.  For example, as we show in Section \ref{sec:Cauchy}, such a lower-than-data dimension MI statistic can be obtained for the Cauchy distribution where a lower-than-data dimension sufficient statistic is a-priori unavailable.  In this manner, MI statistics seamlessly address {\em relevance} to the consumer of the information because it is about some {\em relevant quantity of interest} that MI statistics are maximally informative.  

\section{MI Statistics and the KL Distance  \label{sec:kull}}

Equation (\ref{eq:derivCondx2}) may be rewritten as
\be
\partial_{\bm{r}} \left[ \int P(\bm{q} \mid \bm{x}) log \left( \frac{P(\bm{q} \mid \bm{x})}{P(\bm{q} \mid \bm{r})} \right) d{\bm{q}} \right] \mid_{\bm{r} = {\bm{\xi}_{R}}(\bm{x})} = 0 \label{eq:derivCondx3}
\ee
which, along with the curvature condition, states that 

\begin{quote}
{\bf Theorem 3}. {\em The Kullback-Leibler distance between the posterior distribution conditioned on the statistic and the posterior distribution conditioned on the data is minimized by a MI statistic.}
\end{quote}
Again, note that MI statistics for the quantity of interest are generally not sufficient statistics for the quantity of interest.  Indeed, rather than making the Kullback-Leibler distance zero, as in the case of sufficient statistics, MI statistics are found at local minima of the Kullback-Liebler distance - viewed as a functional of the statistic. This demonstrates how the approach of this paper generalizes that performed by Lindley~\cite{LIND61}.

\section{MI Statistics for the Gaussian distribution \label{sec:Gauss}}

This section details the inference of the one-dimensional MI statistic for the one-dimensional Gaussian distribution.  We take the position parameter of the Gaussian to be $q$, and the the goal is to find $\xi_{R}(\bm{x})$ so that (\ref{eq:derivCondx2}) holds.  From there note that the calculation of $P( q \mid r)$ and $P(q \mid \bm{x})$ is necessary, and by Bayes' theorem therefore it is necessary to find $P( r \mid q)$, which may be written as
\bea
P(r \mid q) & = & \int P( r | q, \bm{x} )  \, P( \bm{x} \mid q)  \, d\bm{x} \nn \\
& = & \int P( r | \bm{x} )  \, P( \bm{x} \mid q)  \, d\bm{x} \nn \\
& = & \int \delta(r - \xi_{R}(\bm{x}))  \, \prod_{i=1}^{N} \frac{e^{-(x_{i}-q)^{2}/2\sigma^{2}}}{\sqrt{2 \pi} \sigma}  \, d\bm{x} \label{eq:rgivenqgauss1}
\eea
The ansatz $\xi_{R}(\bm{x}) = \sum_{i=1}^{N} \lambda_{i} x_{i}$ is useful (and not resrictive since the $\lambda_{i}'s$ are implicitly only restricted to be functions of $\bm{x}$), and making the changes of variables $y_{i} = \lambda_{i} x_{i}$ followed by $u_{i} = \lambda_{i} q$ in (\ref{eq:rgivenqgauss1}) yields a form which may immediately be recognized as the convolution of $N$ Gaussians with means $\mu_{i} = \lambda_{i} q$ and standard deviations $\sigma_{i} = \lambda_{i} \sigma$ respectively,
\be
P(r \mid q)  =  \int \delta(r - \sum_{i=1}^{N} \lambda_{i} q - \sum_{i=1}^{N} u_{i})  \, \prod_{i=1}^{N} \frac{e^{-u_{i}^{2}/2(\sigma \lambda_{i})^{2}}}{\sqrt{2 \pi} (\lambda_{i} \sigma)}  \, d\bm{u}. \label{eq:rgivenqgauss2}
\ee
This has the solution
\be
P(r \mid q) = \phi(0,\sigma')(r-\sum_{i=1}^{N} \lambda_{i} q) \label{eq:rgivenqgauss3}
\ee
where $\sigma' =  \sigma \sqrt{\sum_{i=1}^{N} \lambda_{i}^{2} }$ and $\phi(\cdot,\cdot)(\cdot)$ is the Gaussian density
\be
\phi(\mu,\sigma)(z) = \frac{1}{\sqrt{2 \pi} \sigma} e^{- (z-\mu)^2 / 2 \sigma^2}. \label{eq:1dgaussian}
\ee
Finally, inserting this result into Bayes' theorem with uniform prior to find the posterior distribution of $q$ conditioned on $r$ yields
\be
P(q \mid r) = S \, \phi(0,\sigma')(r-\sum_{i=1}^{N} \lambda_{i} q) \label{eq:qgivenrgauss}
\ee
where $S := \sum_{i=1}^{N} \lambda_{i}$.

The calculation for $P(q \mid \bm{x})$ is similar with the result is that
\be
P(q \mid \bm{x}) = \phi( \overline{\bm{x}} , \frac{\sigma}{\sqrt{N}})(q) \label{eq:qgivenxgauss}
\ee
where $ \overline{\bm{x}} := \frac{1}{N} \sum_{i=1}^{N} x_{i}$.
From the forms of (\ref{eq:qgivenrgauss}) and (\ref{eq:qgivenxgauss}) it is clear that not only will the integrand of (\ref{eq:derivCondx3}) (that equation equivalent to (\ref{eq:derivCondx2})) be minimized, but that it will be zero, if all $\lambda_{i} = 1/N$ is chosen.  This of course is the expected result since $\xi_{R}(x) = \sum_{i=1}^{N} x_{i} / N$ is a sufficient statistic for $q$ when $\sigma$ is known.

Alternatively, to satisfy that the calculation indicated in (\ref{eq:derivCondx2}) is successful at finding the expected result, continue by taking (\ref{eq:qgivenrgauss}) and (\ref{eq:qgivenxgauss}) and substituting them into (\ref{eq:derivCondx2}) to find after some simplification the equation which must be satisfied by $\xi_{R}$
\be
0 = \left[ \int (r-q \sum_{i=1}^{N} \lambda_{i}) \, e^{-(\overline{\bm{x}}-q)^2/2(\sigma/\sqrt{N})^2)} \, dq \right] \mid_{r = \xi_{R}(\bm{x})}.
\ee
This has the unique solution $\xi_{R}(\bm{x}) = \overline{\bm{x}}$ when the arbitrary scale of the inferred statistic is fixed by setting $1 = \sum_{i=1}^{N} \lambda_{i}$.  To conclude this section, the procedure culminating in (\ref{eq:derivCondx3}) or (\ref{eq:derivCondx2}) of finding MI statistics has been shown to produce the expected known result for the Gaussian case.  The next section approaches the Cauchy distribution case for lower than data dimension statistics, where there is no sufficient statistic available and the result is novel.

\section{MI Statistics for the Cauchy Distribution \label{sec:Cauchy}}

This section outlines the inference of the one-dimensional MI statistic for the one dimensional Cauchy distribution.  The detailed steps may be taken similarly to those of the last section but taking the Cauchy distribution instead of the Gaussian distribution. Take the position parameter of the Cauchy to be $q$, and the the goal is to find $\xi_{R}(\bm{x})$ so that (\ref{eq:derivCondx2}) holds. As in the last section it is necessary to determine both $P( q \mid r)$ and $P(q \mid \bm{x})$. Assuming the same ansatz that $\xi_{R}(\bm{x}) = \sum_{i} \lambda_{i} x_{i}$, the necessary convolutions may be carried out with the use of the Fourier convolution theorem, with the results that
\be
P( r \mid q) = \frac{S}{\pi (S^2 + (r-qS)^2) }, \label{eq:rgivenqcauchy}
\ee
\be
P( q \mid r ) = \frac{S}{\pi (S^2 + (r-qS)^2) },  \label{eq:qgivenrcauchy}
\ee
and
\be
P( q \mid \bm{x} ) \propto \prod_{i=1}^{n} \frac{1}{ \pi ( 1 + (x_{i}-q)^2 ) }  \label{eq:qgivenxcauchy}
\ee
where $S := \sum_{i=1}^{N} \lambda_{i}$.   Substituting (\ref{eq:rgivenqcauchy}),   (\ref{eq:qgivenrcauchy}), and (\ref{eq:qgivenxcauchy}) into (\ref{eq:derivCondx2}) yields the equation that must be solved for $\xi_{R}(\bm{x})$
\be
0 = \left[ \int \left(\prod_{i=1}^{n} \frac{1}{ \pi ( 1 + (x_{i}-q)^2 ) }\right) \frac{r/S-q}{1+(r/S-q)^2} \, dq \right] \mid_{r = \xi_{R}(\bm{x})}.
\ee
Rewriting this equation in more suggestive terms, while taking the scale $S=1$, gives the result as an implicit equation for $\xi_{R}(\bm{x})$,
\be
\xi_{R}(\bm{x}) = \int q \, P(q \mid (\bm{x}, \xi_{R}(\bm{x}))\, ) \, dq. \label{eq:xifforcauchy}
\ee
The form of the result (\ref{eq:xifforcauchy}) says that $\xi_{R}(\bm{x})$ is the posterior mean of $q$ given the data {\em and itself as an additional observation}.  This form also suggests that $\xi_{R}(\bm{x})$ could be the posterior mean of $q$ given the data.  However, this is not the case, as a check using the posterior moment forms derived in~\cite{WOLF93} immediately shows.  Further, assuming a value for $\xi_{R}(\bm{x})$ on the right-hand side of (\ref{eq:xifforcauchy}) allows that to be computed in closed form using the results of~\cite{WOLF93}.  This finally yields that the left-hand side is a rational function of the right-hand side, a fixed point equation which may be solved by standard iterative methods.  Other checks immediately show that the solution is not the maximum likelihood solution, nor the median.

To conclude this section, the one-dimensional MI statistic for the Cauchy distribution position parameter has been found as the posterior mean of the position parameter of the Cauchy distribution given the data and the MI statistic, and this statistic is different from the Bayes' estimator which is the posterior mean given the data only.

\section{Approximate MI Inference and Bayes Estimators \label{sec:ApproxAndBayes}}

In many cases of interest, if not in all cases of relevance with high dimensional data, the convolutions that appear similarly to those in (\ref{eq:rgivenqcauchy}) etc. will be quite impossible to do in closed form, and probably in a practical sense will even be numerically intractable.  However, there is an approach that may be taken which does some harm to a fully rigorous Bayesian approach, but which may be necessary.  The idea that is applicable in these cases of difficulty is to directly take $P(\bm{q} \mid \bm{r})$ in (\ref{eq:derivCondx3}) to be Gaussian with parameters $r=\bm{\bm{\xi}_R(\bm{x})}=(\bm{\mu}(\bm{x}), \sigma(\bm{x}))$.
The approximate MI approach just outlined is applied below to finding the approximate MI statistics $(\bm{\mu}(\bm{x}), \sigma(\bm{x}))$.  The approximate MI approach is then contrasted with an alternative approach using the KL distance inverted from that of (\ref{eq:derivCondx3}), one that resembles Maximum Entropy inference.  The rusults of this section hold for any likelihood, as will become apparent.

Take an arbitrary one-dimensional parameterized likelihood parameterized by $q$ (i.e. with $q$ the parameter of interest).  Parameterize the inferred distribution $P(q \mid \bm{r})$ of (\ref{eq:derivCondx3}) as (see (\ref{eq:1dgaussian}))
\be
P(q \mid \bm{r} = (\mu,\sigma) ) = \phi(\mu, \sigma)(q). \label{eq:parameterized}
\ee
Equations (\ref{eq:derivCondx3}) and (\ref{eq:parameterized}) imply that the MI statistic is
\bea
\mu &=& \int q \, P(q \mid \bm{x})  \, dq \nn \\
\sigma^{2} &=& \int (q-\mu)^{2} \, P(q \mid \bm{x}) \, dq
\label{eq:cauchy-kl-norm}
\eea
These quantities are the Bayes' estimators for the mean and standard deviation of the distribution.

If, on the other hand, the inverted form of the KL distance is taken, as it often is in many of the cases we have observed, the statistic $\mu$ is
\be
\mu = 
\frac{\int q \, \phi(\mu, \sigma)(q) \, log\left( P(q \mid \bm{x}) \right) dq}
{ \int \phi(\mu, \sigma)(q) \, log\left( P(q \mid \bm{x}) \right) dq }
\label{eq:cauchy-norm-kl}
\ee
which, along with another non-linear equation for $\sigma$, is a complicated non-linear system to be solved for $\bm{r} = (\mu, \sigma)$.

Note that when the likelihood $P(\bm{x} \mid \bm{q})$ is Gaussian these two approximate approaches produce the same statistic, the posterior mean and standard deviation; but for the Cauchy likelihood, for example, this is not the case, with necessity to solve the complicated nonlinear system. In contrast, the approximate MI inference technique always produces the posterior Bayes' moment estimators.

The difference between the forms of the approximate MI statistics and the inverted-KL statistics appearing in (\ref{eq:cauchy-kl-norm}) and (\ref{eq:cauchy-norm-kl}) respectively makes it clear that one needs a good first-principles approach to the KL distance.

\section{Conclusion}

We have formulated the mutual information based variational principle for statistical inference, a fully Bayesian approach to inference, defined MI statistics for a quantity of interest, shown how the principle may be reformulated as a minimal KL distance principle based on posterior distributions, and demonstrated how inference proceeds, when lower-than-data dimension sufficient statistics are absent, using the Cauchy distribution.  Finally, an approximate approach to the inference of MI statistics was discussed, and the relationship of the resulting statistics to Bayes' estimators and the Maximum Entropy version of the same approximation was noted.

\section{Acknowledgements}

Thanks go to the Data Understanding Group at NASA Ames for their lively and interactive critique, friendship, mentoring, and support. This paper was improved by comments from Dr. Jeremy Frank and Hal Duncan, both of NASA.  Much thanks to Tony O'Hagan for detailed comments.  This work was suported by NASA Center for Excellence in Information Technology contract NAS-214217. This work was supported by NSF grant DMS-98.03756 and Texas ARP grants 003658.452 and 003658.690.

% ----------------------------------------------------------------------------
% BIBLIOGRAPHY
%
%\begin{thebibliography}{500}
%\input{bib.tex}
%\end{thebibliography}
%
%\nocite{*}
\bibliographystyle{plain}
\bibliography{Wolf2}

\end{document}